\documentclass[11pt]{article}
\usepackage{blois}
\usepackage{amssymb,amsmath}

\def\be{\begin{equation}}
\def\ee{\end{equation}}
\def\bea{\begin{eqnarray}}
\def\eea{\end{eqnarray}}

\newcommand{\Photo}{}

\begin{document}

\hfill {\tt CERN-PH-TH/2013-225}
\pagestyle{plain}

\vspace*{4cm}
\title{Indirect search for New Physics: complementarity to direct searches}

\author{ F. Mahmoudi }

\address{Clermont Universit{\'e}, Universit\'e Blaise Pascal, CNRS/IN2P3,\\
LPC, BP 10448, 63000 Clermont-Ferrand, France\\[0.2cm]
CERN Theory Division, Physics Department\\ CH-1211 Geneva 23, Switzerland}

\maketitle\abstracts{
We present an overview on the interplay between direct searches for new physics at the LHC and indirect constraints from the flavour sector, with an emphasis on the implications of the recent LHCb results. The complementarity with the Higgs search results will also be addressed. We show the correlation and complementarity between the different sectors in the context of a few specific examples in supersymmetry.}

\section{Introduction}

While new physics searches are actively pursued by the ATLAS and CMS experiments which aim to detect new particles directly, indirect searches, in particular in the flavour sector, can provide important information and point to specific directions. 

The main focus of beyond the Standard Model (SM) searches at the LHC is Supersymmetry (SUSY). At the end of the 8 TeV run, no signal has been found and searches carried out by the ATLAS and CMS collaborations provided stringent limits on the parameters of the simplest SUSY scenarios. This situation is very uncomfortable since, in spite of the negative search results, it does not allow to exclude SUSY or to point to specific directions. Indeed, one can always argue that if SUSY exists, there is no reason that it should manifest itself in its most simple configurations. Given the large number of parameters involved, the possibilities are numerous and some configurations could simply be experimentally more challenging. Hence, even the current strong exclusion limits leave the door open to low energy supersymmetry. 

On the other hand, indirect searches could provide complementary information that, together with the direct search results, could point to specific and testable scenarios. Moreover, the discovery of a Higgs boson at the LHC and the measurement of its mass and decay rates provide very important information on the SUSY parameters when imposing the SUSY predictions to be in agreement with the measured values. In the following, we will show the complementarity of the direct search results for supersymmetry, information from the Higgs sector, and indirect results from $B$ physics. We discuss such complementarity first in the context of constrained MSSM scenarios and then in a more general unconstrained framework. The interplay with the dark matter sector, although important, will not be covered here.  

\section{$B$ physics observables}

For a long time, the main objective of the $B$ physics experiments was to establish and test the CKM framework. Impressive achievements have been obtained and the CKM paradigm has known a huge success. While such searches are still ongoing, in the recent years the focus of the $B$ physics experiments seems to be changed toward indirect searches for new physics.

Rare decays are amongst the most powerful indirect probes as they are very sensitive to the presence of new particles in the virtual states. The LHCb experiment has a very rich BSM program complemented by $B$ physics searched by ATLAS and CMS. In particular the long awaited $B_s \to \mu^+\mu^-$ decay has been finally observed and several angular observables in the $B \to K^* \mu^+\mu^-$ decay have been measured for the first time. 

The theoretical framework for studying the $B$ physics observables is rather complicated, in particular due to the multi-scale nature of the problem which involves at the same time the new physics scale, the scale of electroweak interactions, QCD interactions and hadronic effects. A solution to this problem is to use the effective field theory approach in which the low and high energy effects are separated using the Operator Product Expansion method. In other words, the heavier degrees of freedom ($t$, $W$, $Z$) are integrated out while the light quarks and gluons are still kept as dynamical particles. This leads to the following effective Hamiltonian~\cite{Buras:1998raa}:
\begin{equation}
\displaystyle {\cal H}_{\rm eff}  =  -\frac{4G_{F}}{\sqrt{2}} V_{tb} V_{ts}^{*} \, \sum_{i=1\cdots10} \Bigl(C_{i}(\mu) \mathcal{O}_i(\mu)+C'_{i}(\mu) \mathcal{O}'_i(\mu)\Bigr)\;,
\end{equation}
where $C_i$ are the Wilson coefficients incorporating physics at short distance which are calculated perturbatively, and $\mathcal{O}_i$ are the local operators representing the long distance part. The primed operators are chirality flipped compared to the non-primed operators, and they are highly suppressed in the SM.
The most relevant operators for rare $B$ decays are:
\begin{align}
\label{physical_basis}
O_1& =  (\bar{s} \gamma_{\mu} T^a P_L c) (\bar{c} \gamma^{\mu} T^a P_L b)\;,\quad
&O_2& = (\bar{s} \gamma_{\mu} P_L c) (\bar{c} \gamma^{\mu} P_L b)\;,\nonumber\\
O_3& =  (\bar{s} \gamma_{\mu} P_L b) \sum_q (\bar{q} \gamma^{\mu} q)\;,\quad
&O_4& = (\bar{s} \gamma_{\mu} T^a P_L b) \sum_q (\bar{q} \gamma^{\mu} T^a q)\;,\nonumber\\
O_5& =  (\bar{s} \gamma_{\mu_1} \gamma_{\mu_2} \gamma_{\mu_3} P_L b) 
                  \sum_q (\bar{q} \gamma^{\mu_1} \gamma^{\mu_2} \gamma^{\mu_3} q)\;,\quad
&O_6& = (\bar{s} \gamma_{\mu_1} \gamma_{\mu_2} \gamma_{\mu_3} T^a P_L b) 
                  \sum_q (\bar{q} \gamma^{\mu_1} \gamma^{\mu_2} \gamma^{\mu_3} T^a q)\;,\nonumber\\
O_7& = \frac{e}{(4\pi)^2} m_b (\overline{s} \sigma^{\mu\nu} P_R b) F_{\mu\nu} \;, \quad
&O_8& = \frac{g}{(4\pi)^2} m_b (\bar{s} \sigma^{\mu \nu} T^a P_R b) G_{\mu \nu}^a \;,
\\ \nonumber
O_9& =  \frac{e^2}{(4\pi)^2} (\overline{s} \gamma^\mu P_L b) (\bar{\ell} \gamma_\mu \ell) \;, \quad
&O_{10}& =  \frac{e^2}{(4\pi)^2} (\overline{s} \gamma^\mu P_L b) (\bar{\ell} \gamma_\mu \gamma_5 \ell)\;. 
\end{align}

This formalism can be extended to New Physics easily, through additional contributions to the Wilson coefficients or additional operators.

The Wilson coefficients are calculated by requiring matching at the $\mu_W$ scale, then evolved  
to the $\mu_b$ scale, which is relevant for $B$ physics calculations, using the renormalisation group equations. To compute the amplitudes, one needs to calculate the hadronic matrix elements which are described in terms of hadronic quantities, {\it i.e.} decay constants and form factors. These quantities are usually the most important source of uncertainty in the calculations.  
Among the most constraining observables are the rare decays $B_s\to\mu^+\mu^-$, $B\to K^*\mu^+\mu^-$, $B \to X_s \gamma$ and leptonic decays such as $B_u \to \tau \nu_\tau$.

The first measurements of the rare decay $B_s\to\mu^+\mu^-$ have been recently announced by the LHCb and CMS collaborations. The combination of their results lead to the branching ratio BR$(B_s\to\mu^+\mu^-) = (2.9 \pm 0.7) \times 10^{-9}$~\cite{Aaij:2013aka,Chatrchyan:2013bka}. 
In terms of Wilson coefficients, this branching ratio is expressed as \cite{Bobeth:2001sq,Mahmoudi:2008tp}:
\begin{eqnarray}
  \label{eq:Bs2mm_formula}
\mbox{BR}(B_s\to\mu^+\mu^-)&=&\frac{G_F^2 \alpha^2}{64\pi^2}f_{B_s}^2
m_{B_s}^3 |V_{tb}V_{ts}^*|^2\tau_{B_s}\sqrt{1-\frac{4m_\mu^2}{m_{B_s}^2}}\\
&&\times\left\{\left(1-\frac{4m_\mu^2}{m_{B_s}^2}\right)
  |C_{Q_1}-C'_{Q_1}|^2+\left|(C_{Q_2}-C'_{Q_2})+2(C_{10}-C'_{10})\frac{m_\mu}{m_{B_s}}\right|^2\right\}\;,\nonumber  
\end{eqnarray}
where\vspace*{-0.6cm}
\begin{align}
Q_1& = \frac{e^2}{(4\pi)^2} (\bar{s} P_R b)(\bar{\ell}\,\ell)\;, \quad
&Q_2& =  \frac{e^2}{(4\pi)^2} (\bar{s} P_R b)(\bar{\ell}\gamma_5 \ell)\;.
\end{align}

In the Standard Model, $C_{Q_1}$ and $C_{Q_2}$ vanish, and $C_{10}$ gets its largest contributions from $Z$ penguin and box diagrams. With the input parameters of~\cite{Mahmoudi:2012un} we obtain BR$(B_s\to\mu^+\mu^-)_{\rm SM}=(3.53\pm 0.38)\times 10^{-9}$. 

The decay $B\to K^*\mu^+\mu^-$ provides a variety of complementary observables
as it gives access to angular distributions in addition to the differential branching fraction. 
The differential decay distribution of the $\bar B \to \bar K ^*(\to K^- \pi^+ ) \ell^+ \ell^-$ decay can be written as a function of three angles $\theta_l$, $\theta_{K^*}$, $\phi$ and the invariant dilepton mass squared ($q^2$) \cite{Kruger:2005ep,Altmannshofer:2008dz}:
\begin{equation}\label{eq:diffAD}
  d^4\Gamma = \frac{9}{32\pi} J(q^2, \theta_l, \theta_{K^*}, \phi)\, dq^2\, d\cos\theta_l\, d\cos\theta_{K^*}\, d\phi \;.
\end{equation}
The angular dependence of $J(q^2, \theta_l, \theta_{K^*}, \phi)$ are then expanded in terms of the angular coefficients $J_i$ which are functions of $q^2$ and can be described in terms of the transversity amplitudes and form factors \cite{Beneke:2001at,Beneke:2004dp}.
Integrating Eq. (\ref{eq:diffAD}) over all angles, the dilepton mass distribution is obtained in terms of the angular coefficients \cite{Altmannshofer:2008dz,Bobeth:2008ij}:
\begin{equation}
\frac{d\Gamma}{dq^2} = \frac{3}{4} \bigg( J_1 - \frac{J_2}{3} \bigg)\;.
\label{eq:dBR}
\end{equation}
The forward-backward asymmetry $A_{FB}$, which benefits from reduced theoretical uncertainty, is defined as:
\begin{equation}
A_{\rm FB}(q^2)  \equiv
     \left[\int_{-1}^0 - \int_{0}^1 \right] d\cos\theta_l\, 
          \frac{d^2\Gamma}{dq^2 \, d\cos\theta_l} \Bigg/\frac{d\Gamma}{dq^2}
          =  -\frac{3}{8} J_6 \Bigg/ \frac{d\Gamma}{dq^2}\;.
\label{eq:AFB}
\end{equation}
Another clean observable is the zero--crossing of the forward-backward asymmetry ($q_0^2$) for which the form factors cancel out at leading order. $q_0^2$ depends on the relative sign of $C_7$ and $C_9$ and its measurement enables to remove the sign ambiguity. 

The longitudinal polarisation fraction $F_L$ can also be constructed as the ratio of the transversity amplitudes and contains less theoretical uncertainty from the form factors. It reads:
\begin{equation}
 F_L(s) = \frac{-J_2^c}{d\Gamma / dq^2}\;.
\end{equation}
The SM predictions and experimental values for these observables are given in Table~\ref{tab:experiment}.
\begin{table}[!t]
\begin{center}
\begin{tabular}{|l|l|l|l|l|l|l|}\hline 
  Observable                                                                & SM prediction & Experiment       \\ \hline \hline
  $10^7 \mbox{GeV}^2 \times \langle dBR/dq^2\; (B \to K^* \mu^+ \mu^-) \rangle_{[1,6]}$ & $0.47 \pm 0.27 $        & $0.42 \pm 0.04 \pm 0.04$   \\ \hline
  $\langle A_{FB}(B \to K^* \mu^+ \mu^-) \rangle_{[1,6]}$         & $-0.06 \pm 0.05 $        & $-0.18 ^{+0.06+0.01}_{-0.06-0.01}$   \\ \hline
  $q_0^2 (B \to K^* \mu^+ \mu^-)/\mbox{GeV}^2$      & $4.26 ^{+0.36}_{-0.34} $        & $4.9  ^{+1.1}_{-1.3}$   \\ \hline
  $\langle F_{L}(B \to K^* \mu^+ \mu^-) \rangle_{[1,6]}$          & $0.71 \pm 0.13 $        & $0.66 ^{+0.06+0.04}_{-0.06-0.03}$   \\ \hline
 \end{tabular}
\caption{SM predictions and experimental values of $B\to K^*\mu^+\mu^-$ observables~{\protect\cite{Mahmoudi:2012un}}.
\label{tab:experiment}}
\end{center}
\end{table}
The decay $B\to K^*\mu^+\mu^-$ gives access to many other angular observables which will be measured in the near future~\footnote{New results for more optimised observables have been released after this presentation by LHCb~\cite{Aaij:2013qta}.}.

The decay $B \to X_s \gamma$ provides also important constraints on new physics scenarios. It proceeds through electromagnetic penguin loops, involving $W$ boson in the Standard Model, in addition to charged Higgs boson, chargino, neutralino and gluino loops in supersymmetric models. The branching ratio of $B \to X_s \gamma$ can be written as \cite{Misiak:2006zs}
\begin{equation}
{\rm BR}(\bar{B} \to X_s \gamma)= \mathrm{BR}(\bar{B} \to X_c e \bar{\nu})_{\rm exp} \left| \frac{ V^*_{ts} V_{tb}}{V_{cb}} \right|^2 \frac{6 \alpha}{\pi C} \left[ P(E_0) + N(E_0) + \epsilon_{em} \right] \;,
\end{equation}
with $C = | V_{ub}/V_{cb} |^2 \times \Gamma[\bar{B} \to X_c e \bar{\nu}]/\Gamma[\bar{B} \to X_u e \bar{\nu}]$. $P(E_0)$ and $N(E_0)$ denote respectively the perturbative and non perturbative contributions, which involve the Wilson coefficients $C_{1-8}$, with $E_0$ a cut on the photon energy. $\epsilon_{em}$ is an electromagnetic correction. The calculation is performed at NNLO accuracy in the SM and 2HDM and at NLO (partial NNLO) in SUSY \cite{Bobeth:2004jz,Misiak:2006ab,Hermann:2012fc}. With the latest PDG input parameters we obtain ${\rm BR}(\bar{B} \to X_s \gamma)_{\rm SM}=(3.08 \pm 0.23) \times 10^{-4}$ which can be compared to the world average experimental value ${\rm BR}(\bar{B} \to X_s \gamma)_{\rm exp}=(3.43 \pm 0.21 \pm 0.07)\times 10^{-4}$~\cite{Amhis:2012bh}. 

The purely leptonic decay $B_u \to \tau \nu_\tau$ on the other hand occurs via $W^+$ and $H^+$ mediated annihilation processes already at tree level. This decay is helicity suppressed in the SM, but there is no such suppression for the charged Higgs exchange, and at high $\tan\beta$ the two contributions can be of similar magnitudes. This decay is thus very sensitive to the charged Higgs boson properties and provides important constraints.
The branching ratio of $B_u \to \tau \nu_\tau$ reads \cite{Hou:1992sy,Akeroyd:2003zr,Itoh:2004ye}
\begin{equation}
{\rm BR}(B_u\to\tau\nu_\tau)=\frac{G_F^2f_B^2|V_{ub}|^2}{8\pi}\tau_B m_B m_\tau^2\left(1-\frac{m_\tau^2}{m_B^2}\right)^2 \left[1-\left(\frac{m_B^2}{m_{H^+}^2}\right)\frac{\tan^2\beta}{1+\epsilon_0\tan\beta}\right]^2 \;,
\end{equation}
where $\epsilon_0$ corresponds to a two loop SUSY correction. Using $|V_{ub}|=(4.15\pm0.49)\times10^{-3}$ and $f_B = 194 \pm 10$ MeV, the SM branching ratio amounts to BR($B_u\to\tau\nu_\tau)_{\rm SM}=(1.15 \pm 0.29)\times 10^{-4}$ which is similar to the combination of the most recent Belle and Babar results BR($B_u\to\tau\nu_\tau)_{\rm exp}=(1.14 \pm 0.23)\times 10^{-4}$~\cite{Adachi:2012mm,Lees:2012ju}.

\section{Interplay with direct searches}

To illustrate the constraining power of the flavour observables and the complementarity with direct searches for new physics, we consider in the following first a rather simple MSSM scenario, CMSSM, where the universality assumptions at the GUT scale allow us to reduce the number of free parameters to a handful, and next a more general framework, the pMSSM, where no universality assumption is imposed. The SUSY spectra are generated with {\tt SOFTSUSY}~\cite{Allanach:2001kg} and flavour observables are calculated with {\tt SuperIso}~\cite{Mahmoudi:2008tp,Mahmoudi:2007vz}.

\subsection{Constrained MSSM}

\begin{figure}[!t]
\begin{center}
\includegraphics[width=7cm]{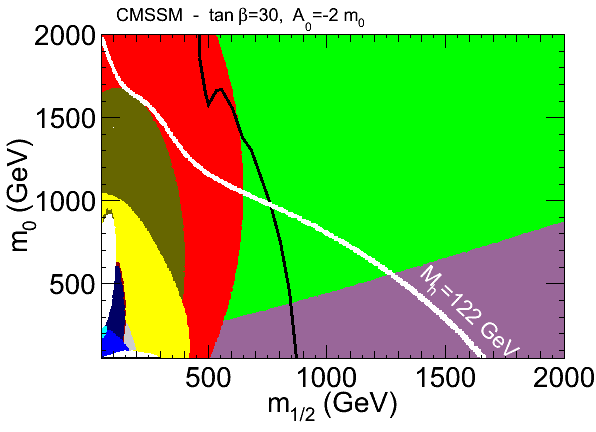}\includegraphics[width=7cm]{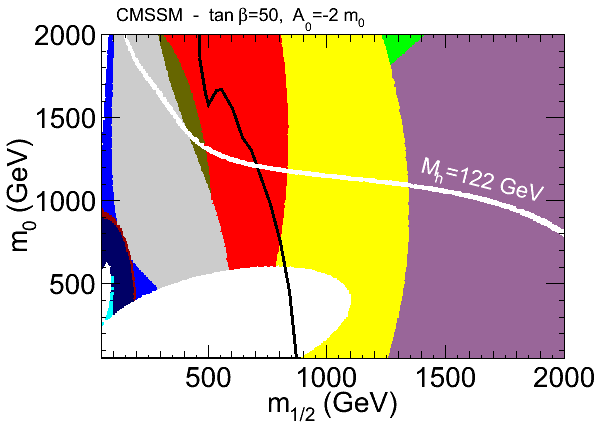}~\raisebox{0.4cm}{\includegraphics[width=2cm]{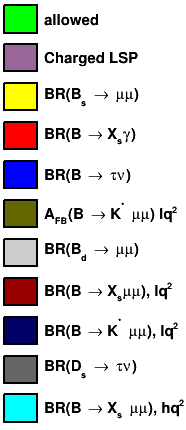}}
\end{center}\vspace*{-0.2cm}
\caption{Flavour constraints in the CMSSM, in the $(m_{1/2} ,m_0)$ parameter plane with $A_0 = -2m_0$, for $\tan\beta = 30$ in the left and $\tan\beta = 50$ in the right. The black lines delimit the ATLAS SUSY direct search limits with 20.3~fb$^{-1}$ of data and the white lines show where the Higgs mass can reach a value of 122 GeV.\label{fig:cmssm1}}
\end{figure}
We study the effects of imposing flavour constraints on the CMSSM parameters by performing
flat scans varying the CMSSM parameters in the ranges: $m_0, m_{1/2} \in [50,3000] {\;\rm GeV};\;\tan\beta \in [1,60];\;  A_0 \in [-10,10] {\;\rm TeV};\; {\rm sign}(\mu)>0$.
A comparison between different flavour observables in the plane $(m_{1/2} ,m_0)$ is given in Fig.~\ref{fig:cmssm1}, where the constraints from flavour observables described in the previous section are shown for $\tan\beta = 30$ and 50. The latest ATLAS SUSY direct search limit with 20.3~fb$^{-1}$ of data \cite{ATLAS-CONF-2013-047} in the same plane is superimposed for comparison. As can be seen in CMSSM at large  $\tan\beta$, the flavour constraints are stronger than direct searches for SUSY partners.

\begin{figure}[!t]
\begin{center}
\includegraphics[width=6cm]{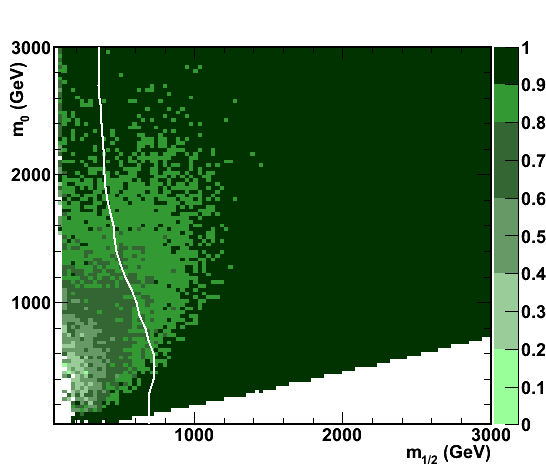}\quad\quad\includegraphics[width=6cm]{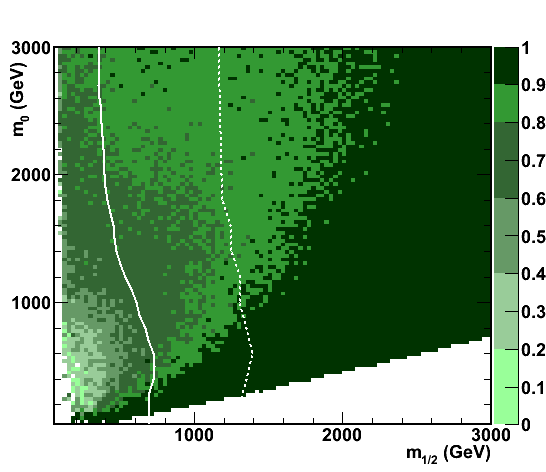}
\end{center}
\caption{Fraction of CMSSM points compatible with the current (left) and ultimate (right) 95\% C.L. constraints on BR($B_s \to \mu^+ \mu^-$) in the $(m_{1/2},m_0)$ parameter plane~{\protect\cite{Arbey:2012ax}}. The continuous line shows the region excluded by searches at 8 TeV with 5.8~fb$^{-1}$ of data and the dotted line the reach estimated at 14~TeV with 300~fb$^{-1}$.
\label{fig:cmssm2}}
\end{figure}

Figure~\ref{fig:cmssm2} shows the fraction of CMSSM points compatible with the current  measurement of BR($B_s \to \mu^+ \mu^-$) and the expected ultimate precision with an accuracy of 5\% in the $(m_{1/2},m_0)$ plane when all the parameters, including $\tan\beta$, are varied. They are compared to the region excluded at 95\% C.L. by the SUSY searches in channels with missing transverse energy with 5.8~fb$^{-1}$ of data at 8~TeV~\cite{ATLAS:2012ona} and the expected reach with 300~fb$^{-1}$ at 14~TeV~\cite{Abdullin:1998pm}, which shows that the sensitivity through the $B_s \to \mu^+ \mu^-$ decay improves approximately as the reach of direct searches. However, while searches in the jets + MET channels are directly sensitive to the $m_{1/2}$ and $m_0$ parameters, the $B_s \to \mu^+ \mu^-$ decay probes a complementary region of the CMSSM parameter space, accessible to direct searches only through the $H/A \rightarrow \tau^+ \tau^-$ channel. 

\begin{figure}[t!]
\begin{center}
\includegraphics[width=6.cm]{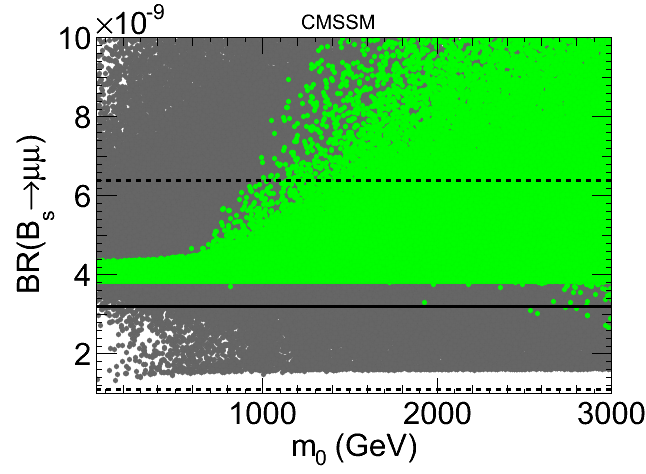}%
\includegraphics[width=6.cm]{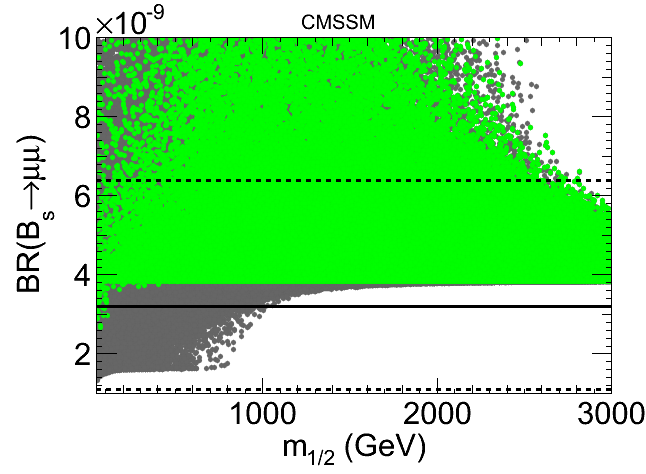}\\[0.cm]
\includegraphics[width=6.cm]{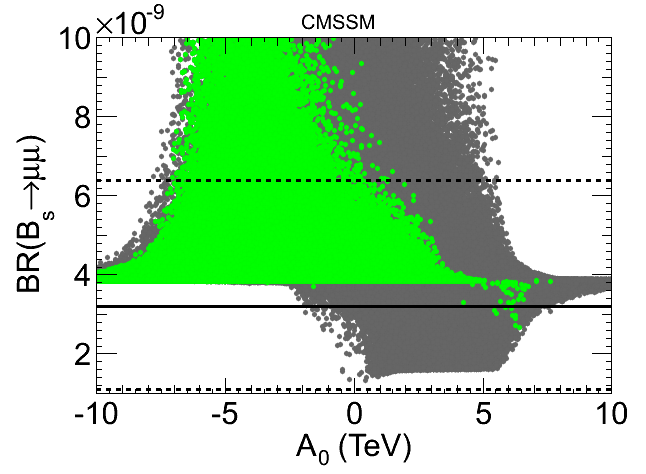}%
\includegraphics[width=6.cm]{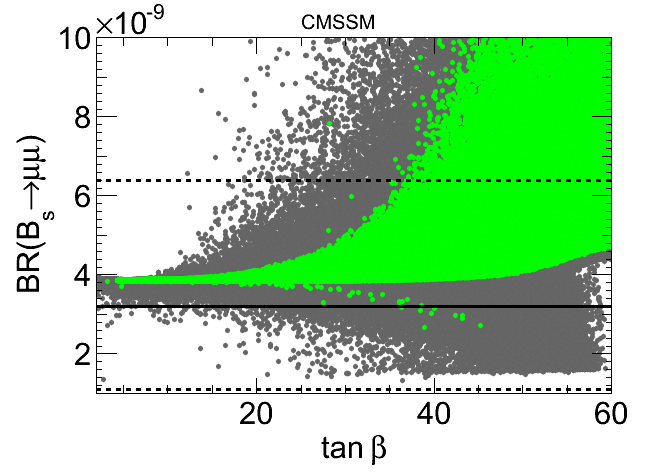}%
\caption{BR($B_s \to \mu^+ \mu^-$) vs.\ CMSSM parameters $m_0$ (upper left), $m_{1/2}$ (upper right), $A_0$ (lower left), $\tan\beta$ (lower right)~{\protect\cite{Arbey:2012ax}}. The solid lines correspond to the central value of the BR($B_s \to \mu^+ \mu^-$) measurement, and the dashed lines to the 2$\sigma$ experimental deviations. The green points are those in agreement with the Higgs mass constraint.
}
\label{fig:cmssm3}
\end{center}
\end{figure}

In Fig.~\ref{fig:cmssm3}, we show the BR($B_s \to \mu^+ \mu^-$) values as functions of the four CMSSM parameters, comparing all the generated points to those consistent with the lightest Higgs boson $h$ mass range, $123 < M_h < 129$~GeV~\cite{Aad:2012gk,Chatrchyan:2012gu}.
Branching fraction values below $\sim 3\times10^{-9}$ can be reached for $m_{1/2} \lesssim 1$ TeV, $0 \lesssim A_0 \lesssim 6$ TeV and $\tan\beta \gtrsim 20$. However, once the Higgs mass limits are imposed, the allowed points have the BR($B_s \to \mu^+ \mu^-)$ at values which are equal to, or larger than, the SM prediction. 
As a consequence, in the CMSSM, it is not possible to have BR($B_s \to \mu^+ \mu^-)$ smaller than the SM prediction and at the same time be in agreement with the SUSY and Higgs search results. 

\subsection{Phenomenological MSSM}

\begin{figure}[t!]
\begin{center}
\includegraphics[width=5.3cm]{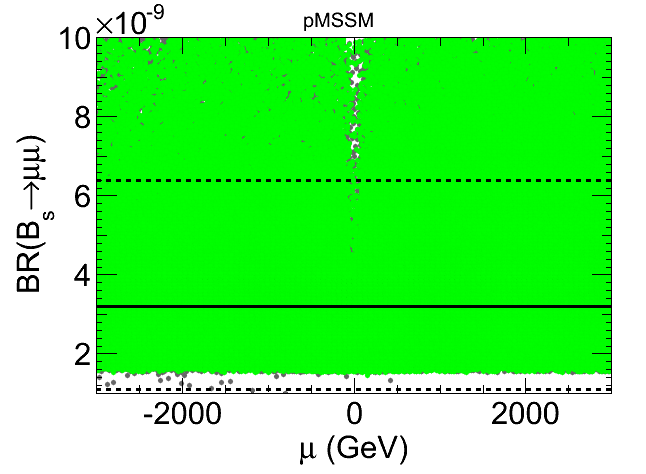}%
\includegraphics[width=5.3cm]{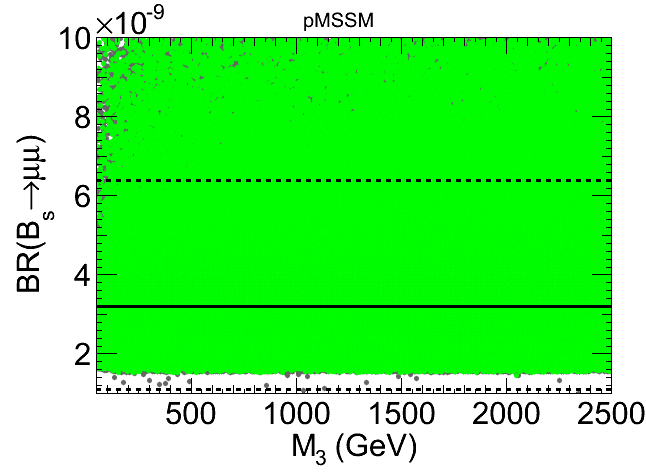}%
\includegraphics[width=5.3cm]{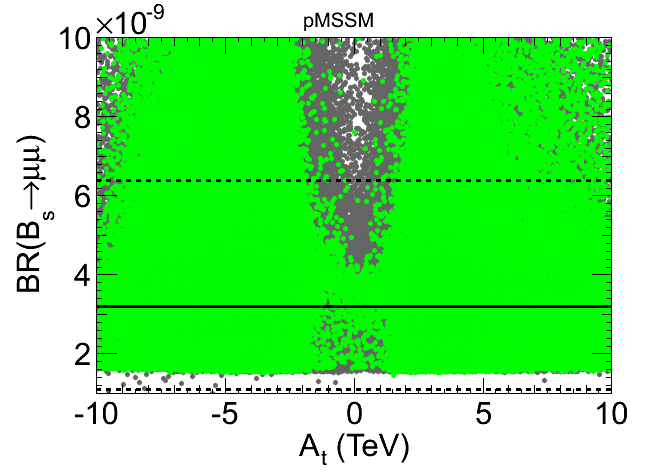}\\[0.cm]
\includegraphics[width=5.3cm]{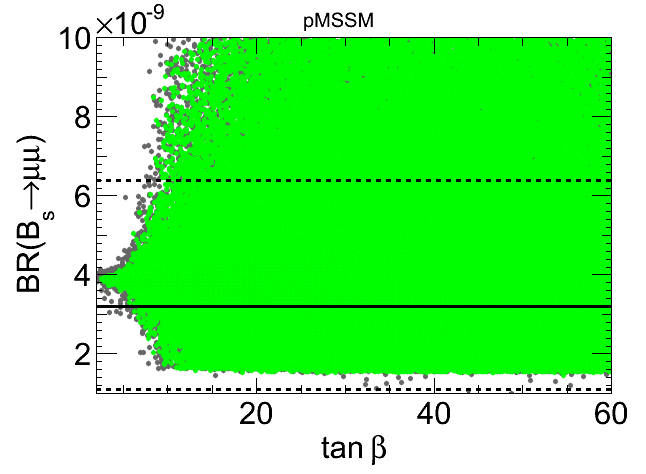}%
\includegraphics[width=5.3cm]{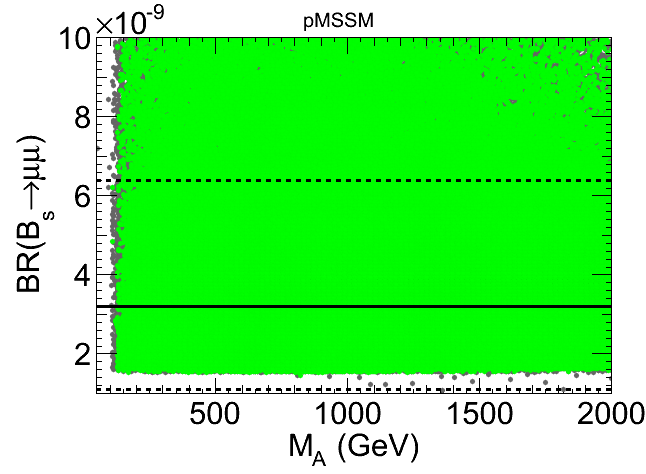}%
\includegraphics[width=5.3cm]{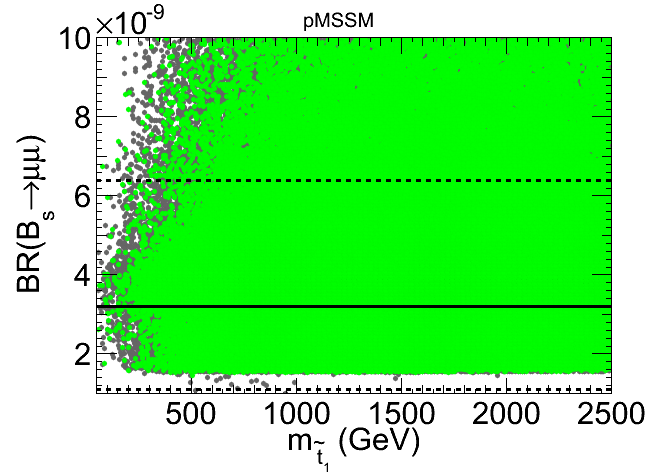}%
\caption{BR($B_s \to \mu^+ \mu^-$) vs.\ parameters $\mu$ (upper left), $M_3$ (upper central), $A_t$ (upper right), $\tan\beta$ (lower left), $M_A$ (lower central) and $m_{\tilde t_1}$ (lower right)~{\protect\cite{Arbey:2012ax}}. The solid lines correspond to the central value of the BR($B_s \to \mu^+ \mu^-$) measurement, and the dashed lines to the 2$\sigma$ experimental deviations. The green points are those in agreement with the Higgs mass constraint. 
}
\label{fig:pmssm}
\end{center}
\end{figure}

\begin{figure}[t!]
\begin{center}
\includegraphics[width=7.cm]{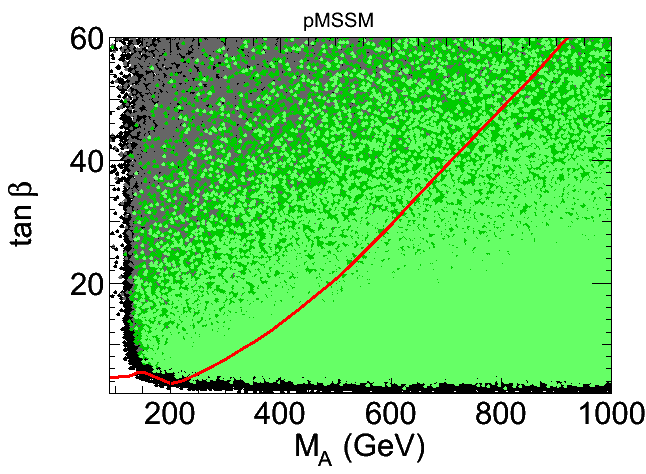}
\includegraphics[width=7.cm]{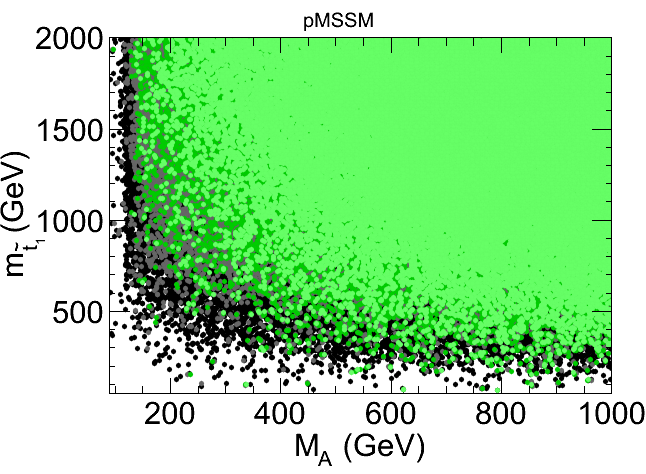}
\caption{Constraints from BR($B_s \to \mu^+ \mu^-$) in the ($M_A,\tan\beta$) and ($M_A,m_{\tilde{t}_1}$) parameter planes~{\protect\cite{Arbey:2012ax}}. The black points corresponds to all the valid pMSSM points and those in grey to the points for which $123 < M_h < 129$ GeV. The dark green points in addition are in agreement with the current BR($B_s \to \mu^+ \mu^-$) range, while the light green points are in agreement with the prospective LHCb BR($B_s \to \mu^+ \mu^-$) range. The red line indicates the region excluded at 95\% C.L. by the CMS $A/H\to\tau^+\tau^-$ searches with 17~fb$^{-1}$ of data.
}
\label{fig:pmssm3}
\end{center}
\end{figure}

The pMSSM relaxes the correlations introduced by the mass universality assumptions of the CMSSM and allows us to study the influence of the flavour observables on the MSSM parameters in a more general set-up. Since only a few of these parameters enter in the calculation of the $B_s \to \mu^+ \mu^-$ branching fraction, the pMSSM offers also a viable framework to study the complementarity of the constraints from this process with those derived from direct searches by ATLAS and CMS.
The analysis presented here adopts the method and tools described in \cite{Arbey:2011un,Arbey:2011aa}. We perform flat scans over the 19 pMSSM parameters in the ranges:
\begin{equation}
\begin{gathered} 
 M_1,M_2 \in [-2500,2500] {\;\rm GeV};\; M_3 \in [50,2500] {\;\rm GeV};\; \tan\beta \in [1,60]\\
M_A \in [50,2000] {\;\rm GeV};\; A_t,A_b,A_\tau \in [-10,10] {\;\rm TeV};\; \mu \in [-3,3] {\;\rm TeV}\\
m_{\tilde{\ell},\tilde{\tau}_L}, m_{\tilde{\ell},\tilde{\tau}_R} \in [50,2500] {\;\rm GeV};\; m_{\tilde{U},\tilde{D},\tilde{b},\tilde{t}_R},m_{\tilde{Q},\tilde{Q_3}_L} \in [50,3500] {\;\rm GeV}\;.
\end{gathered}
\end{equation}
The dependence of the BR($B_s \to \mu^+ \mu^-$) values calculated at each pMSSM point with the most relevant pMSSM parameters is given in Fig.~\ref{fig:pmssm} for all the valid points and those having $123 < M_h < 129$ GeV. Contrary to the case of the CMSSM, here even after imposing the Higgs mass constraints a sizeable number of points with a value of BR($B_s \to \mu^+ \mu^-$) below the SM prediction (down to $0.5 \times 10^{-9}$) is obtained. These low values are reached for $\tan\beta \gtrsim 10$ and $m_{\tilde{t}_1} \gtrsim 300$ GeV.
The impact of the present and future determinations of BR($B_s \to \mu^+ \mu^-$) on the parameters most sensitive to its rate, ($M_A,\tan\beta$) and ($M_A,m_{\tilde{t}_1}$), is shown in Fig.~\ref{fig:pmssm3} where we give all the valid pMSSM points from our scan, those with $123 < M_h < 129$ GeV and, highlighted in green, those in agreement with the present BR($B_s \to \mu^+ \mu^-$) range and the ultimate constraint at 95\% C.L. (with 5\% accuracy).
The constraints from BR($B_s \to \mu^+ \mu^-$) affect the same pMSSM region, at large values of $\tan\beta$ and small values of $M_A$, as also probed by the dark matter direct detection constraints and, more importantly, the $H/A \to \tau^+ \tau^-$ direct Higgs searches at the LHC~\cite{CMS-HIG-2012-050,Aad:2011rv}. The search for the $H/A \to \tau^+ \tau^-$ decay has already excluded a significant portion of the parameter space where large effects on BR($B_s \to \mu^+ \mu^-$) are expected. We also note that the stop sector is further constrained by direct searches in $b$-jets + MET channels, which disfavour small values of $m_{\tilde{t}_1}$. The figure shows that it is difficult for $M_A$ and $m_{\tilde{t}_1}$ to be simultaneously light. This is yet another example showing the additional information that can be obtained by combining flavour and collider constraints.

Let us now have a closer look at the interplay with Higgs sector. For simplicity, we consider the decoupling ({\it i.e.} $M_A^2 \gg M_Z^2$) and large $\tan \beta$ regime. The Higgs mass at tree level can be written as
\begin{equation}
M_{h}^2 \approx M_Z^2 \cos^2 2\beta \left[1-\frac{M_Z^2}{M_A^2} \sin^2 2\beta\right]\;,
\end{equation}
which leads to $M_h^2 \le M_Z^2$ if it were not for the radiative corrections which push the Higgs mass upward. The dominant one-loop contribution arises from top and stop loops~\cite{Okada:1990vk,Ellis:1990nz,Haber:1990aw}:
\begin{equation}
(\Delta M_h^2)_{\tilde t} \approx  \frac{3 \sqrt{2} G_F}{2\pi^2}\, m_t^4 \left [ -\log\left(\frac{m_t^2}{m_{\tilde t}^2}\right)+ \frac{X_t^2}{m_{\tilde t}^2} \left ( 1 - \frac{X_t^2}{12 m_{\tilde t}^2} \right )  \right ]\;,
\end{equation}
with $X_t = A_t -\mu / \tan\beta$ and $M_S = \sqrt{m_{\tilde t_1} m_{\tilde t_2}}$.
The important parameters for the Higgs mass are therefore the stop mass, $\mu$, $A_t$ and $\tan \beta$.
These parameters are also relevant for the Higgs decay rates. For example, the diphoton channel receives contributions from stop, sbottom, stau, charged Higgs boson and chargino loops, which can be parametrised as~\cite{Haisch:2012re}:
\begin{equation}
\kappa_\gamma \equiv \frac{\Gamma (h \to \gamma\gamma)_{\rm MSSM}}{\Gamma (h \to \gamma\gamma)_{\rm SM}}\approx  \frac{1}{F_W- \frac{4}{3}} \, \Bigg[  -\frac{4}{3} \, \kappa_{\tilde t}  -\frac{1}{3} \, \kappa_{\tilde b} - \kappa_{\tilde \tau}  + \kappa_{H^\pm} +  \kappa_{\chi^\pm} \Bigg]\;.
\end{equation}
One way to eventually enhance the diphoton rate would be through the stau loop if the staus are light enough~\cite{Carena:2011aa}. The stau contribution can be written as 
\begin{equation}
\kappa_{\tilde \tau}  \approx -\frac{m_\tau^2 X_\tau^2}{4 m_{\tilde \tau_1}^2 m_{\tilde \tau_2}^2}\;,
\end{equation}
where $X_\tau = A_\tau - \mu  \tan\beta$.
Enhancement of $h\to\gamma\gamma$ could be possible in particular for small $m_{\tilde \tau}$, large $\mu$ and large $\tan\beta$.

It is remarkable that the flavour observables, and in particular BR($B \to X_s \gamma$) and BR($B_s \to \mu^+ \mu^-$) are also dependent on the same parameters. To show this feature, taking into account only the most important corrections due to the Wilson coefficient $C_7$ (for illustration purpose), we can parametrise the $B \to X_s \gamma$ branching ratio normalised by the SM expression as~\cite{Haisch:2012re}:
\begin{equation}
R_{X_s} = \frac{{\rm BR} (B \to X_s \gamma)_{\rm MSSM}}{{\rm BR} (B \to X_s \gamma)_{\rm SM}} \approx 1- 2.61 \,  \Delta C_7 + 1.66 \, ( \Delta C_7 )^2\;,
\end{equation}
with contributions from charged Higgs and chargino loops:
\begin{equation}
\Delta C_7^{H^\pm} \approx \frac{m_t^2}{3 M_{H^\pm}^2} \left (\ln \frac{m_t^2}{M_{H^\pm}^2} + \frac{3}{4} \right ) \,,\qquad \Delta C_7^{\chi^\pm} \approx - \mu A_t \, \tan\beta \, \frac{m_t^2}{M_S^4} \, g (x_{\tilde t \mu})\;,
\end{equation}
with $x_{\tilde t \mu} = M_{S}^2/\mu^2$ and 
$g (x) = -\frac{7 x^2-13 x^3}{12 \left (1-x \right )^3}- \frac{2 x^2 - 2 x^3 -3 x^4}{6 \left (1-x \right )^4}\, \ln x$.
As can be seen, the stop mass, charged Higgs mass (hence $M_A$), $\mu$, $A_t$ and $\tan \beta$ are also important for $B \to X_s \gamma$ branching ratio.

\begin{figure}[t!]
\begin{center}
\includegraphics[width=7.cm]{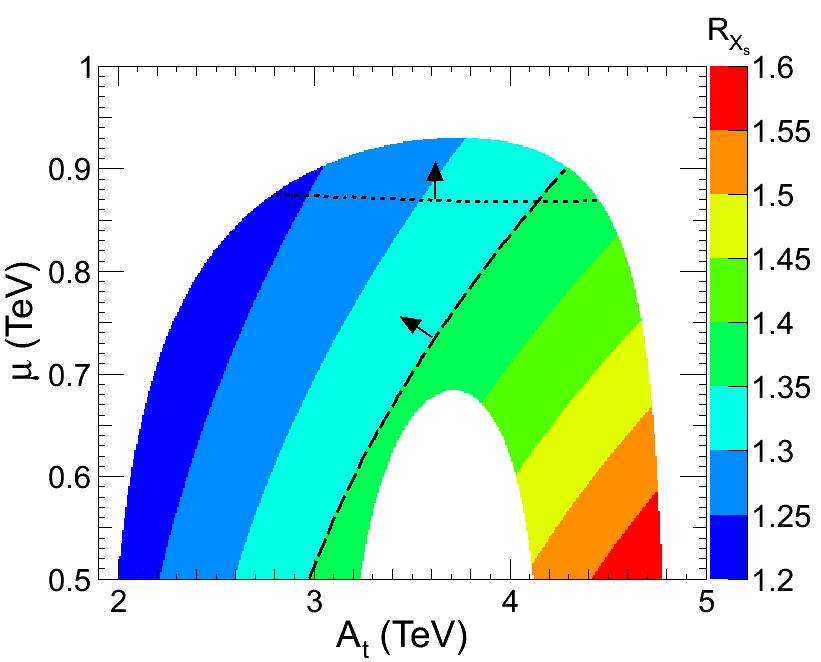}\includegraphics[width=7.cm]{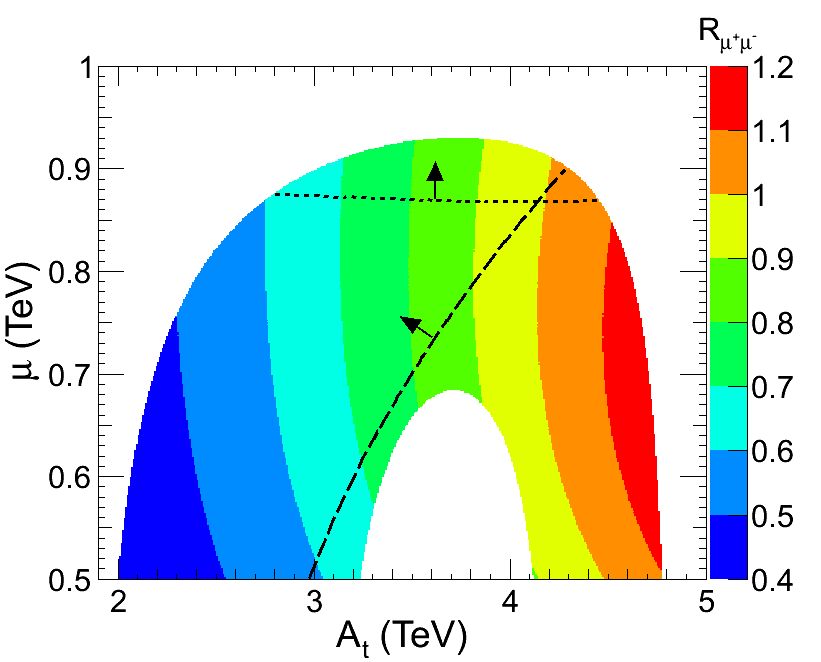}%
\caption{Predictions for BR($B \to X_s \gamma$) normalised to the SM value, $R_{X_s}$ in the left and BR($B_s\to\mu^+\mu^-$) normalised to SM, $R_{\mu^+ \mu^-}$ in the right~{\protect\cite{Haisch:2012re}}. The coloured region shows the area where $121 < M_h < 129$ GeV is satisfied. The dotted black lines indicate the parameter regions with $h \to \gamma\gamma$ above the SM value, while the dashed black lines delimit the $95\% \,{\rm CL}$ regions favoured by $B \to X_s \gamma$.}
\label{fig:light-stau}
\end{center}
\end{figure}

Similar correlations can also be seen in the $B_s \to \mu^+ \mu^-$ branching ratio, which can be parametrised as~\cite{Haisch:2012re}:
\begin{equation}
R_{\mu^+\mu^-} = \frac{{\rm BR} (B_s \to \mu^+ \mu^-)_{\rm MSSM}}{{\rm BR} (B_s \to \mu^+ \mu^-)_{\rm SM}} \approx  1 - 13.2  \; C_P + 43.6 \left (C_{Q_1}^2 + C_{Q_2}^2 \right )\;,
\end{equation}
where the dominant contribution to the $C_{Q_1}$ and $C_{Q_2}$ can be written as
\begin{equation}
C_{Q_1} \approx -C_{Q_2} \approx -\mu A_t \, \frac{\tan^3\beta}{(1+\epsilon_b \, \tan\beta)^2} \; \frac{m_t^2}{M_S^2} \, \frac{m_b  m_\mu}{4\sin^2\theta_W M_W^2 M_A^2} \, f (x_{\tilde t \mu})\;,
\end{equation}
with $f(x) =-\frac{x}{1-x} - \frac{x}{(1-x)^2} \, \ln x$. Again, the dependence on the $m_{\tilde{t}}$, , $M_A$, $\mu$, $A_t$ and $\tan\beta$ is manifest.
One can therefore expect important correlations between the flavour observables and the Higgs sector. An example of such correlations, in a scenario with light stau particles is displayed in Fig.~\ref{fig:light-stau}.
Here we impose the constraint $121 < M_h < 129$ GeV, which is satisfied in the coloured region of the plots. In the left plot, the colour scale corresponds to BR($B \to X_s \gamma$) normalised to the SM value, and in the right plot, it corresponds to BR($B_s\to\mu^+\mu^-$) normalised to SM. The Higgs to diphoton signal strength larger than 1 is only achieved in the upper part of the plot (above the dotted lines). This region has already started to be probed as it corresponds to a part of the parameter space where BR($B_s\to\mu^+\mu^-$) is smaller than the SM, once the $B \to X_s \gamma$ constraints are applied. These correlations are striking and could be tested in the near future and hence may become very valuable as guidelines and consistency checks.

\begin{figure}[t!]
\begin{center}
\includegraphics[width=6.cm]{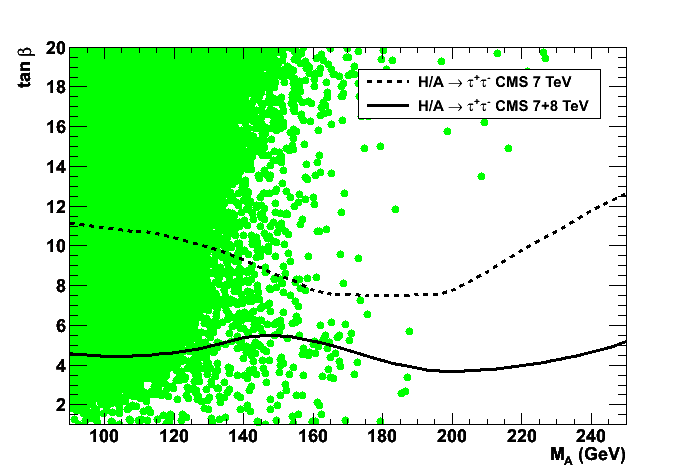}%
\includegraphics[width=6.cm]{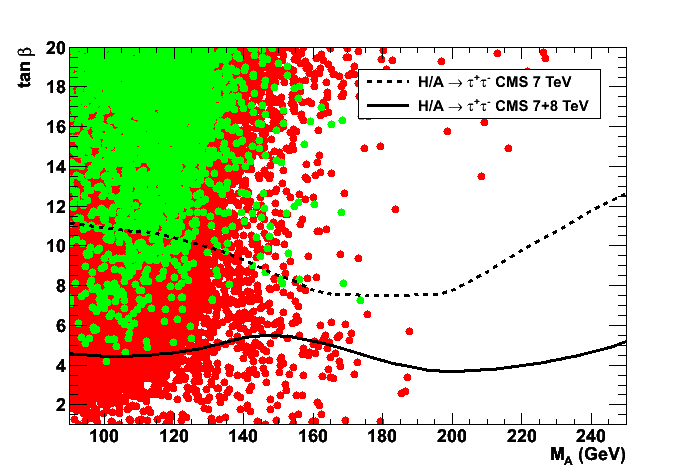}\\[0.cm]
\includegraphics[width=6.cm]{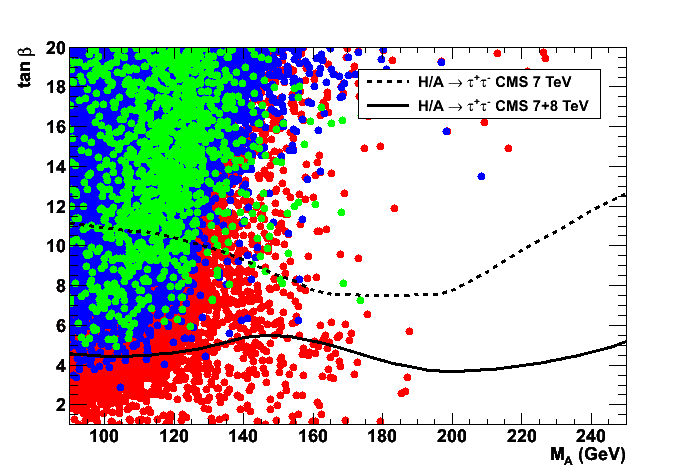}%
\includegraphics[width=6.cm]{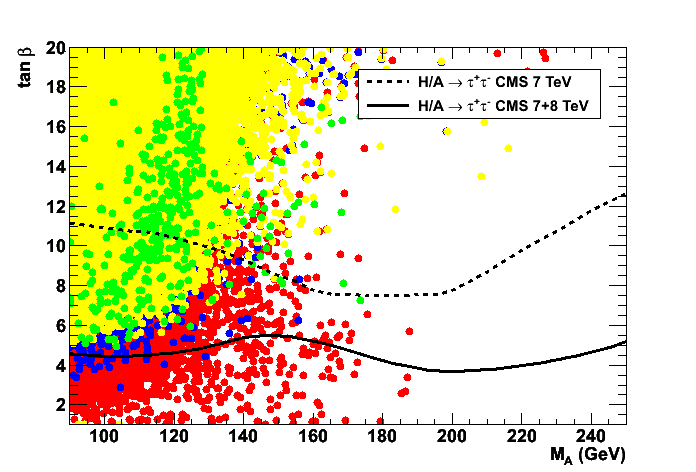}%
\caption{Parameter plane ($M_A,\tan\beta$) with points for the heavier $H$ boson to be observed with a mass in the interval $121 < M_H < 129$ GeV (green points, upper left panel)~{\protect\cite{Arbey:2012bp}}. The red points are excluded by BR($B \to X_s \gamma$) (upper right), the blue by BR($B_u\to\tau\nu_\tau$) (lower left) and the yellow by BR($B_s\to\mu^+\mu^-$) (lower right). The CMS excluded regions from the 2011 and 2012 $H/A \to \tau^+ \tau^-$ searches are shown by the dashed and continuous lines, respectively.}
\label{fig:pmssm4}
\end{center}
\end{figure}
It has been advocated that the Higgs boson discovered at the LHC could correspond to the MSSM heavy Higgs $H$ state. In Fig.~\ref{fig:pmssm4}, we draw the points with $121 < M_H < 129$ GeV in the $(M_A,\tan\beta)$ parameter plane~\cite{Arbey:2012bp}. While the $H/A \to \tau^+ \tau^-$ searches set strong limits in the $(M_A,\tan\beta)$ parameter plane, we can see that a substantial part of the points satisfying the heavy Higgs hypothesis still survives at low $\tan\beta$. However, applying the constraints from BR($B_s\to\mu^+\mu^-$), BR($B\to X_s \gamma$) and BR($B_u\to\tau\nu_\tau$) rules out this possibility, as can be seen in the figure. This again shows the important information that can be obtained when taking advantage of the data from both direct and indirect searches.

\section{Conclusions}

The interplay between direct and indirect searches for new physics has entered a new era with the start of the LHC. Striking correlations exist between in particular $B$ physics sector and direct Higgs and SUSY searches at the LHC. Exploiting such complementarity is the only way to squeeze the parameter spaces and point to specific scenarios. Moreover in the fortunate case of new particle discovery in the next LHC run, consistency checks would be essential for a deeper understanding of the underlying physics. Examples of such testable cross checks and correlations in supersymmetry have been presented here.

\section*{Acknowledgments}

I would like to thank the organisers of the Rencontres de Blois 2013 for the great organisation of the conference and for inviting me to give this talk.


\end{document}